\documentclass[showpacs]{revtex4}
\usepackage{indentfirst}
\usepackage[mathscr]{eucal}
\usepackage[dvips]{graphicx}
\usepackage{amsmath}
\def\lsim{\mathrel{\raise2pt\hbox to 8pt{\raise -5pt\hbox{$\sim$}\hss{$<$}}}}
\def\bmath#1{\mbox{\boldmath $#1$}}
\def\ds{\displaystyle}

\begin{document}
\title{Role of the Coulomb and the vector-isovector $\rho$ potentials in the isospin asymmetry of nuclear pseudospin}
\author{R. Lisboa and M. Malheiro}
\affiliation{Instituto de F{\'\i}sica, Universidade Federal Fluminense,
24210-340 Niter\'oi, Brazil}
\author{P. Alberto}
\affiliation{Departamento de F{\'\i}sica and Centro de F{\'\i}sica
Computacional, Universidade de Coimbra, P-3004-516 Coimbra,
Portugal} 
\date{\today}


\begin{abstract}
\noindent We investigate the role of the Coulomb and the
vector-isovector $\rho$ potentials in the asymmetry of the neutron
and proton pseudospin splittings in nuclei. To this end, we solve
the Dirac equation for the nucleons using central vector and
scalar potentials with Woods-Saxon shape and $Z$ and $N-Z$
dependent Coulomb and $\rho$ potentials added to the vector
potential. We study the effect of these potentials on the energy
splittings of proton and neutron pseudospin partners along a Sn
isotopic chain. We use an energy decomposition proposed in a
previous work to assess the effect of a pseudospin-orbit potential
on those splittings. We conclude that the effect of the Coulomb
potential is quite small and the  $\rho$ potential gives the main
contribution to the observed isospin asymmetry of the pseudospin
splittings. This isospin asymmetry results from a cancellation of
the various energy terms and cannot be attributed only to the
pseudospin-orbit term, confirming the dynamical character
of this symmetry pointed out in previous works.
\end{abstract}
\maketitle


\section{Introduction}

 Pseudospin was introduced in the late 1960s \cite{kth,aa}
to account for the
quasidegeneracy of single-nucleon states with quantum numbers
($n$, $\ell$, $j= \ell + 1/2$) and
($n-1$, $\ell+2$, $j= \ell$ + 3/2)
where $n$, $\ell$, and
$j$ are the radial, the orbital, and the total angular momentum
quantum numbers, respectively. These levels have the
same ``pseudo-orbital'' angular momentum quantum number,
$\tilde{\ell} = \ell + 1$, and ``pseudospin'' quantum number,
$\tilde s = 1/2$.

 The pseudospin symmetry has been analyzed in nonrelativistic models
by several works \cite{bahri,mosh,mosk}. The subject was revived
in recent years when Ginocchio \cite{gino} presented a
relativistic interpretation for this symmetry showing that the
quantum number $\tilde\ell$ is the orbital quantum number of the
lower component of the Dirac spinor for spherical potentials.
Moreover, he showed that $\tilde\ell$ is a good quantum number in
a relativistic theory for the nucleon with scalar $S$ and vector
$V$ potentials with opposite signs and the same magnitude, i.e.,
$\Sigma=S+V=0$. Actually, this condition can be relaxed to demand
that only the derivative of $\Sigma$ is zero \cite{arima}, but, if
$\Sigma$ goes to zero at infinity, both conditions are equivalent.
These findings are in agreement with the nuclear phenomenology
obtained in relativistic mean field (RMF) theories, in view of the
fact that the scalar and vector potentials in RMF for the nucleus
\cite{wal,zim,delf,delf1}, although large, cancel each other to a
great extent, giving a relatively small negative value for
$\Sigma$ on the nucleon mass scale \cite{furn0}. The structure of
the radial nodes occurring in pseudospin levels as well as the
nuclei wave function for them were also explained in recent works
\cite{gino2,gino5,levi1}. Unfortunately, because in the
nonrelativistic reduction of the Dirac equation $\Sigma$ acts as
the binding potential, it cannot be set to zero in nuclei, since
in that case there would not be any bound state.

 So the question remains how the pseudospin symmetry is realized in
nuclei, namely how some levels of pseudospin partners become
degenerate or near degenerate. A partial answer to this question
has been given in two previous works \cite{pmmdm_prl,pmmdm_prc},
in which were shown that, using Woods-Saxon potentials, this
degeneracy not only depends on the value of the depth of the
$\Sigma$ potential but also on its characteristic extension
(radius) and its surface steepness (diffusivity). Furthermore,
using an energy decomposition based on a Schr\"{o}dinger-like
equation for the radial lower component of the Dirac spinor, it
was shown that the near-degeneracy of some pseudospin doublets
arises from a cancellation between the terms in that
decomposition, and that the pseudospin-orbit term is in general
bigger than the splitting itself. This agreed with the findings of
Refs.~\cite{marcos,marcos2} in which the pseudospin-orbit coupling was
shown to be a nonperturbative quantity. The conclusion was that
pseudospin symmetry in nuclei has a dynamical character.

 The systematics found in Refs.~\cite{pmmdm_prl,pmmdm_prc} for the
pseudospin dependence on the shape of the nuclear mean fields
allowed the explanation for the difference between the energy
splittings of proton and neutron pseudospin partners. This was
done by revealing the influence of the $\rho$ potential $V_\rho$,
 which appears in RMF theories, on pseudospin-orbit interaction,
favoring pseudospin symmetry for neutron levels as is found
experimentally. However, the role of the Coulomb potential $V_{\rm
Coul}$ for the proton splittings, which has the opposite sign of
$V_\rho$, was not considered. A previous study of the isospin
dependence of the pseudospin symmetry was done in Ref.~\cite{meng0} for
the Zr and Sn isotopes, where it was shown that the pseudospin
symmetry becomes better for exotic nuclei with a highly diffuse
potential.

 In this paper we examine in a more quantitative way this isospin
asymmetry, by including explicitly the Coulomb and the $V_\rho$
potentials in the nuclear mean fields for the Sn isotopic chain
and assess the role of each potential in that asymmetry
\cite{ronai}. We will show that even though the pseudospin orbit
term is very sensitive to the Coulomb potential $V_{\rm Coul}$, the
net effect of this potential in the proton pseudospin energy
splitting is quite small. The main contribution to the asymmetry
of the neutron and proton pseudospin splitting in nuclei comes
from the vector-isovector $V_{\rho}$ potential confirming the
results found in Refs.~\cite{pmmdm_prl,pmmdm_prc}.

 The paper is organized as follows. In Sec.~II we review briefly
the formalism of the Dirac equation with scalar and vector
potentials and the energy decomposition based on a
Schr\"{o}dinger-like equation for the lower component of the Dirac
spinor. In Sec.~III the neutron and proton mean field
potentials in a Sn isotopic chain are parametrized by Woods-Saxon
form as functions of $A$, $N$ and $Z$ including explicitly the
$V_{\rho}$ and $V_{\rm Coul}$ potentials. The variation of the
pseudospin splittings with varying $A$ and how Coulomb and $\rho$
potentials affect those splittings, separating the contribution
of each term in the energy decomposition, are presented in Sec.~IV. 
In Sec.~V we give a summary of the work and draw the
conclusions.


\section{Dirac equation and pseudospin symmetry}

 The Dirac equation for a particle of
mass $m$ in an external scalar potential $S$ and vector potential $V$ is
given by
\begin{equation}
H\Psi=[\bmath{\alpha}\cdot\bmath{p}\,+\,\beta(m\,+\,S)\,+\,V]\Psi\,=\,\epsilon\,\Psi\,,\label{dirac}
\end{equation}
where ${\bmath\alpha}$ and $\beta $ are the usual Dirac matrices.
Defining $\Delta=V-S$ and the upper and lower components as
$\Psi_\pm=[(1 \pm \beta)/2]\Psi$, we have
\begin{eqnarray}
\bmath{\alpha}\cdot\bmath{p}\,\Psi_+&=&(E+2m-\Delta)\Psi_-\label{dirac_first1}\\
\bmath{\alpha}\cdot\bmath{p}\,\Psi_-&=&(E-\Sigma)\Psi_+\,,\label{dirac_first2}
\end{eqnarray}
where $E=\epsilon-m$ is the binding energy. The Hamiltonian in
Eq.~(\ref{dirac}) is invariant under SU(2) transformations when $S
= V$ or $S = - V$ \cite{smith,bell,levi}.  The second case
corresponds to the realization of pseudospin symmetry. As referred
before, this symmetry is related to the orbital quantum number of
the lower component of the Dirac spinor $\Psi_-$. This can be seen
if one writes the corresponding decoupled second-order equations
of Eqs.~(\ref{dirac_first1}) and (\ref{dirac_first2}), assuming
that the potentials $\Sigma$ and $\Delta$ are radial functions:
\begin{eqnarray}
p^2\Psi_{+}&=&-\frac{\Delta'}{E+2m-\Delta}\bigg(-\frac{\partial\hfill}{\partial{r}}+\frac{1}{r}\bmath{\sigma}\cdot\bmath{L}\bigg)\Psi_{+}+(E+2m-\Delta)(E-\Sigma)\Psi_{+}
\label{psiplus}\\
p^2\Psi_{-}&=&-\frac{\Sigma'}{E-\Sigma}\bigg(-\frac{\partial\hfill}{\partial{r}}+\frac{1}{r}\bmath{\sigma}\cdot\bmath{L}\bigg)\Psi_{-}+(E+2m-\Delta)(E-\Sigma)\Psi_{-}\,,
\label{psiminus}
\end{eqnarray}
where the primes denote derivatives with respect to $r$ and
$\bmath{\sigma}$ is the usual diagonal $4\times4$ spin matrix. The
spinors $\Psi_\pm$ can be separated into radial and angular parts,
$\Psi_+={\rm i}\,G_\kappa(r)\,\Phi^+_{\kappa\,m_j}(\theta,\phi)$
and $\Psi_-=-F_\kappa(r)\,\Phi^-_{\kappa\,m_j}(\theta,\phi)$,
where $\kappa$ is the
 quantum number related to spin-orbit coupling and $m_j$ is the eigenvalue of the
$J_z$ operator. Doing this separation in Eqs.~(\ref{psiplus}) and
(\ref{psiminus}) we arrive at the radial equations
\begin{eqnarray}
\frac{1}{r^2}\frac{d\hfill}{d\,r}\bigg(r^2\,\frac{d\,G_\kappa}{d\,r}\bigg)-
\frac{\ell(\ell+1)}{r^2}G_\kappa+\frac{\Delta'}{E+2m-\Delta}\bigg(G'_\kappa+
\frac{1+\kappa}{r}G_\kappa\bigg)+
(E+2m-\Delta)(E-\Sigma)G_\kappa=0\\\label{upper}
\frac{1}{r^2}\frac{d\hfill}{d\,r}\bigg(r^2\,\frac{d\,F_\kappa}{d\,r}\bigg)-
\frac{\tilde\ell(\tilde\ell+1)}{r^2}F_\kappa+\frac{\Sigma'}{E-\Sigma}\bigg(F'_\kappa
+\frac{1-\kappa}{r}F_\kappa\bigg)+(E+2m-\Delta)(E-\Sigma)F_\kappa=0\,,\label{lower}
\end{eqnarray}
since one has
$\bmath{\sigma}\cdot\bmath{L}\Phi^\pm_{\kappa\,m_j}=-(1\pm\kappa)\Phi^\pm_{\kappa\,m_j}$
\cite{palberto}. The term with $1-\kappa$ in Eq.~(\ref{lower}) is
the pseudospin-orbit term. From that equation one sees that,
should it be possible to set $\Sigma'=0$, $\tilde\ell$ would be a
good quantum number. Since the sign of $\kappa$ determines whether
one has parallel or antiparallel spin
\begin{equation}
\kappa=\left\{\begin{array}{cl}
                    -(\ell+1)         &\quad\,j   =  \ell + \frac{1}{2}\\
                      \ell            &\quad\,j   =  \ell - \frac{1}{2}
                   \end{array}\right.\,,
\end{equation}
and $\tilde\ell=\ell-\kappa/|\kappa|$, one sees that pairs of
states with $\kappa=-(\ell+1)$ and $\kappa=\ell+2$ have the same
$\tilde\ell=\ell+1$, the quantum numbers of the pseudospin
partners.

The second-order differential equation for $\Psi_-$
[Eq.~(\ref{psiminus})] can be written as a Schr\"{o}dinger-type equation
\cite{pmmdm_prc}:
\begin{equation}
\frac{p^2}{2m^*}\Psi_{-}+\frac{1}{2\,m^*}\frac{\Sigma'}{E-\Sigma}\bigg(-\frac{\partial\hfill}{\partial{r}}+\frac{1}{r}\bmath{\sigma}\cdot\bmath{L}\bigg)\Psi_-\,+\Sigma\,\Psi_-=E\,\Psi_-\,.\label{lower_schr}
\end{equation}
where $m^*=(E+2m-\Delta)/2$ is an energy- and $r$-dependent
effective mass. This equation allows an energy decomposition to be
performed by taking the expectation value of each term relative to
the spinor $\Psi_-$
\begin{equation}
\bigg\langle\,\frac{p^2}{2m^*}\bigg\rangle+\big\langle\,V_{\rm{PSO}}\big\rangle+
\big\langle\,V_{\rm{D}}\big\rangle+\langle\Sigma\rangle=E\,,\label{aveg_schroed}
\end{equation}
where
\begin{eqnarray}
\label{kinetic} \bigg\langle
\frac{p^2}{2m^*}\bigg\rangle&=&\frac{\displaystyle \int^\infty_0
dr\, \frac{1}{\displaystyle 2m^*} \bigg[-F_{\kappa}\frac{
d\hfill}{d r}\bigg(r^2\,\frac {d F_{\kappa}}{
d r}\bigg)+ \tilde\ell(\tilde\ell+1)F_{\kappa}^2\bigg]}
{\displaystyle\int_0^\infty r^2\,dr\,F_{\kappa}^2}\,,\\
\label{pso} \big\langle V_{\rm PSO}\big\rangle&=&{\rm P}\,
\frac{\displaystyle \int^\infty_0 r\,dr\, \,\frac{1}{2
m^*}\frac{\Sigma'}{E-\Sigma}(\kappa-1)F_{\kappa}^2}
{\displaystyle\int_0^\infty r^2\,dr\,F_{\kappa}^2}\,,\\
\label{darwin} \big\langle V_{\rm D}\big\rangle&=&-{\rm P}\,
\frac{\displaystyle \int^\infty_0 r^2\,dr\, \,\frac{1}{2
m^*}\frac{\Sigma'}{E-\Sigma}F_{\kappa}
\frac { d F_{\kappa}}{d r}}
{\displaystyle\int_0^\infty r^2\,dr\,F_{\kappa}^2}\,,\\
\label{sigma_aver} \langle\Sigma\rangle&=&\frac{\displaystyle
\int^\infty_0 r^2\,\,dr\,\Sigma F_{\kappa}^2}
{\displaystyle\int_0^\infty r^2\,dr\, F_{\kappa}^2\,}\ .
\end{eqnarray}
In these expressions, P stands for the principal value of the
integral. In Ref.~\cite{pmmdm_prc} it was shown that the
pseudospin-orbit and Darwin terms ($\big\langle V_{\rm PSO}\big\rangle$ and
$\big\langle V_{\rm D}\big\rangle$, respectively) obey a sum rule
that provides a numerical check for the computation of the
principal values. The decomposition (\ref{aveg_schroed}) will
allow us to evaluate the importance of each term, and in
particular of $\big\langle V_{\rm PSO}\big\rangle$, for the
pseudospin splitting calculations shown in Sec.~IV.


\section{PARAMETRIZATION OF THE NUCLEAR POTENTIAL FOR THE {Sn} ISOTOPIC CHAIN}

 In RMF theories, the main contributions to the isospin breaking in nuclei are given
by the Lorentz vector electrostatic (Coulomb) potential and the
isovector-vector potential $V_\rho$ coming from the interaction of
$\rho$ mesons with the nucleons \cite{chiappa}. These
potentials modify the vector potential to
\begin{equation}
V=\left\{\begin{array}{ll}
V_c+V_\rho+V_{\rm Coul} &\quad \hbox{for protons}\\
V_c-V_\rho &\quad \hbox{for neutrons,} \end{array}\right.
\end{equation}
where $V_c$ stands for the central (isoscalar) vector potential.
Using the definition of $\Sigma$ and $\Delta$ potentials we see
that both are affected in the same way; i.e.,
\begin{equation}
\label{sigmadelta}
\Sigma \hbox{ or } \Delta=\left\{\begin{array}{ll}
( \Sigma_c \hbox { or } \Delta_c )+V_\rho+V_{\rm Coul} &\quad \hbox{for protons}\\
( \Sigma_c \hbox { or } \Delta_c )-V_\rho &\quad \hbox{for neutrons,} \end{array}\right.
\end{equation}
where $\Sigma_{c}=S+V_{c}$ and $\Delta_{c}=V_{c}-S$. As in
Ref.~\cite{pmmdm_prl,pmmdm_prc}, we use a Woods-Saxon form
\begin{equation}\label{Eq:W-S}
U(r)=\frac{U_{0}}{1+\exp[(r-R)/a]}
\end{equation}
for $\Sigma_c$ and $\Delta_c$ potentials with depth $U_0$, radius $R$,
and diffusivity $a$ to be determined later for each potential.

For $V_\rho$, we again use a Woods-Saxon shape,
\begin{equation}\label{Vrho}
V_\rho(r)=\frac{V_{o\rho}}{ 1+\exp[(r-R_\rho)/a_\rho]}\ ;
\end{equation}
for $V_{\rm Coul}$ we take the proton electrostatic potential
energy in a uniform spherical charge distribution of radius $R_c$,
\begin{eqnarray}
\label{VCoul}
V_{\rm Coul}(r)=\left\{\begin{array}{ll}\ds
            \frac{1}{4\pi\epsilon_{0}}\frac{Ze^2}{2R_{C}}
\biggl(3-\frac{r^2}{R^2_{C}}\biggr)&,\quad\;r<\;R_{C}\\[0.3cm]
          \ds  \frac{1}{4\pi\epsilon_{0}}\frac{Ze^2}{r}&,\quad\;r\geq\;R_{C}.
            \end{array}
       \right.
\end{eqnarray}

\begin{figure}[!h]
\parbox[!t]{8.5cm}{
\begin{center}
\includegraphics[width=8.5cm,height=8.5cm]{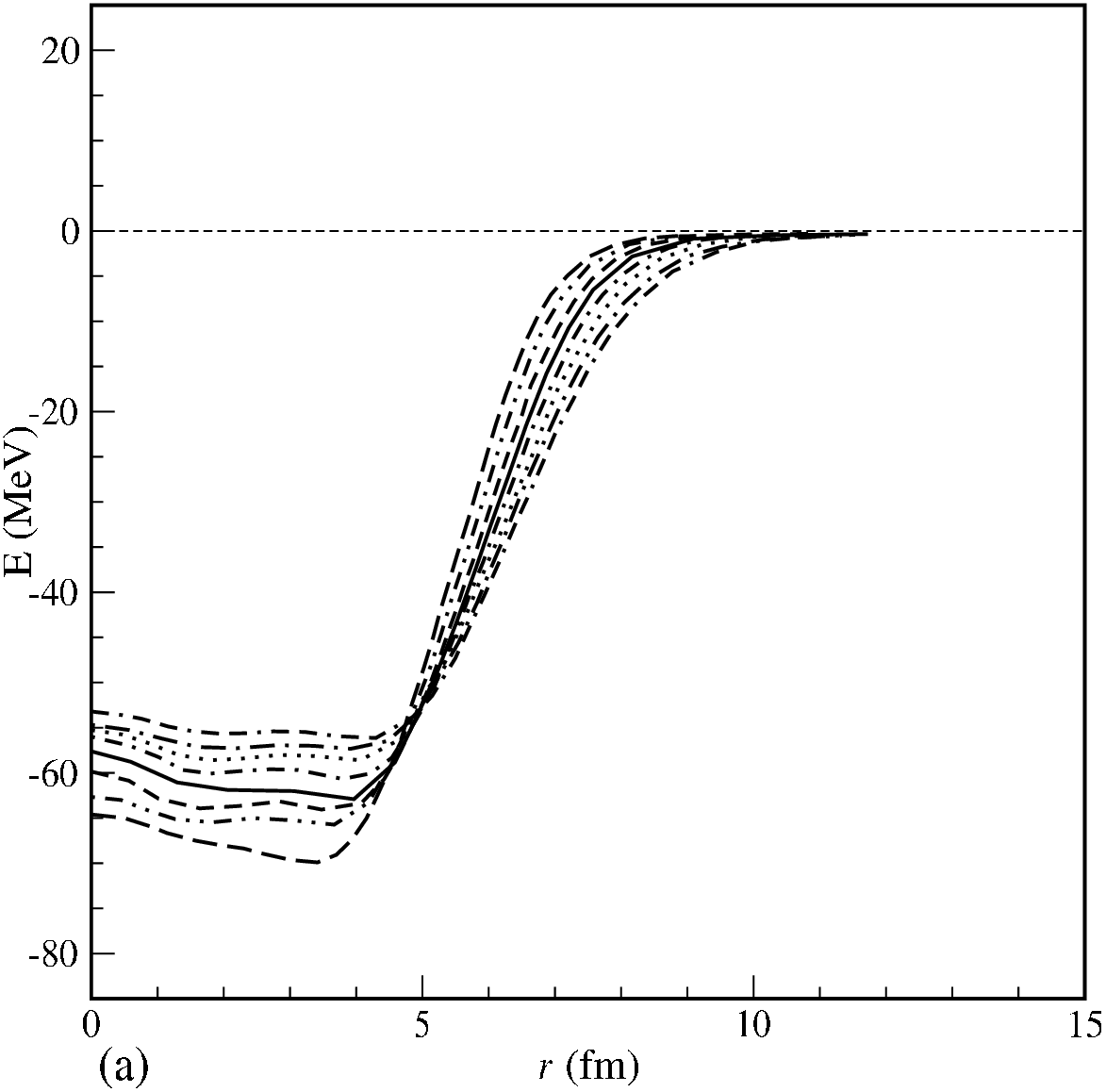}
\end{center}\vspace*{-0.4cm}}
\hfill
\parbox[!h]{8.5cm}{
\begin{center}
\includegraphics[width=8.5cm,height=8.5cm]{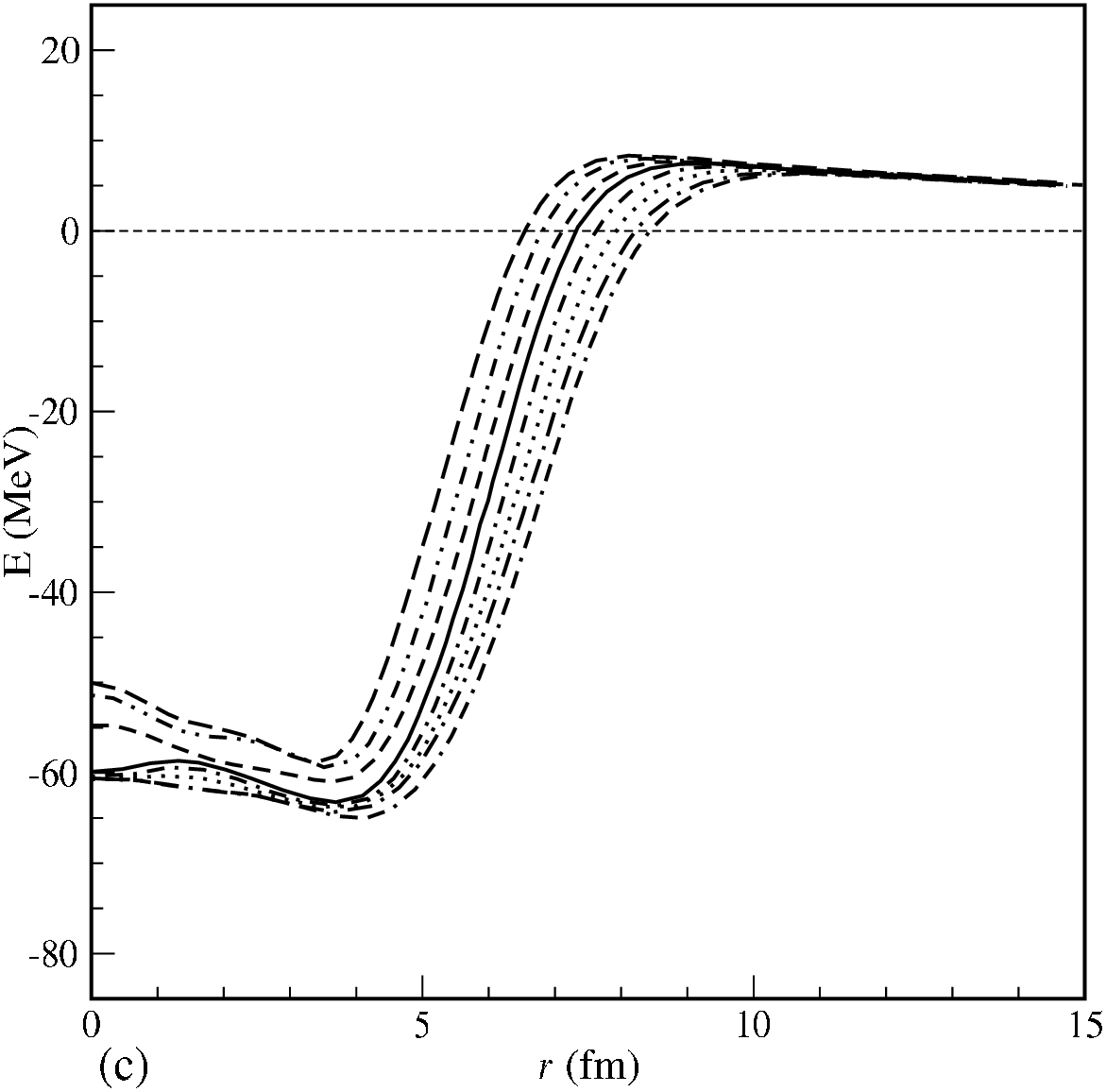}
\end{center}\vspace*{-0.4cm}}
\hfill
\parbox[!h]{8.5cm}{
\begin{center}
\includegraphics[width=8.5cm,height=8.5cm]{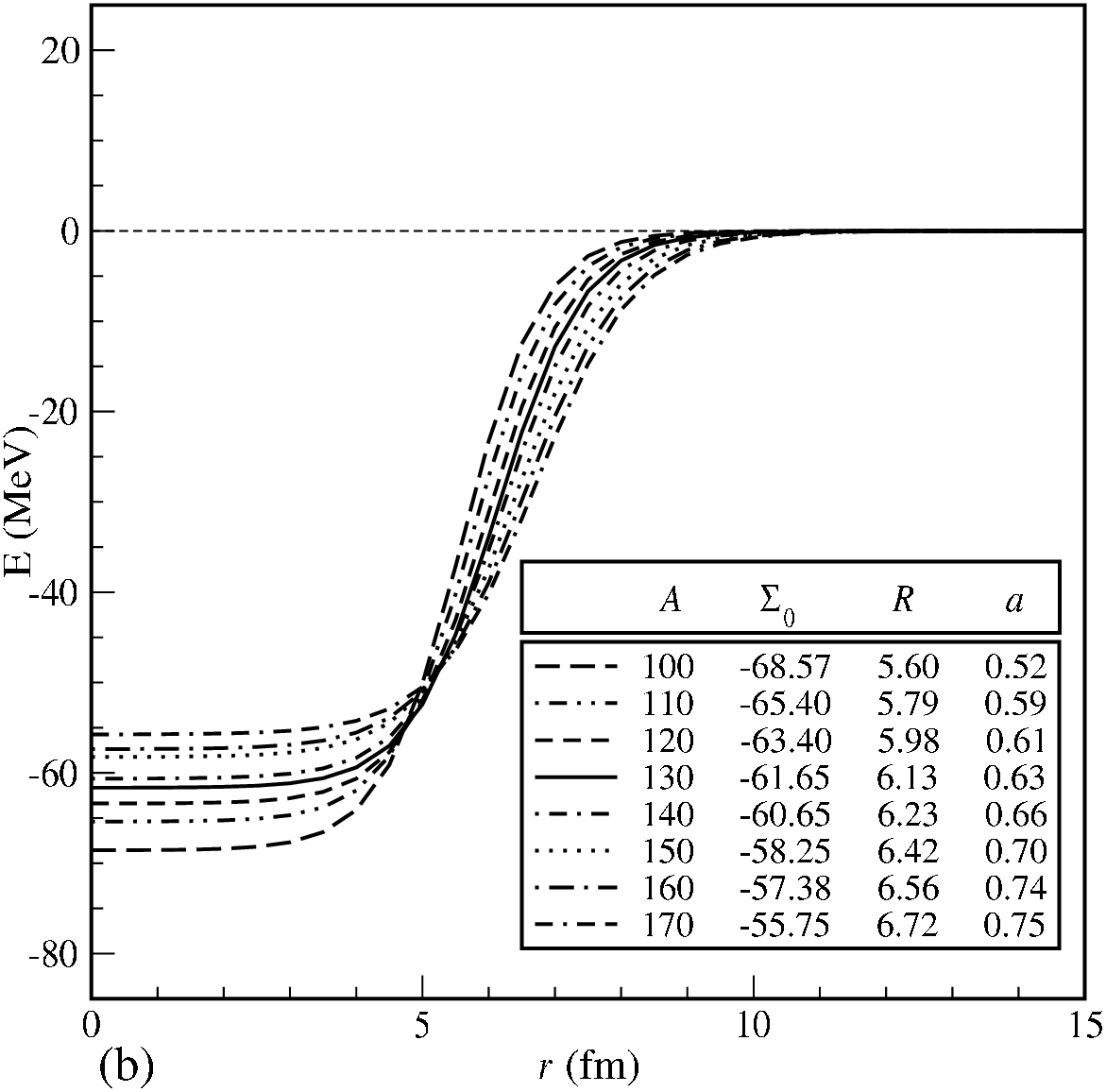}
\end{center}\vspace*{-0.4cm}}
\hfill
\parbox[!h]{8.5cm}{
\begin{center}
\includegraphics[width=8.5cm,height=8.5cm]{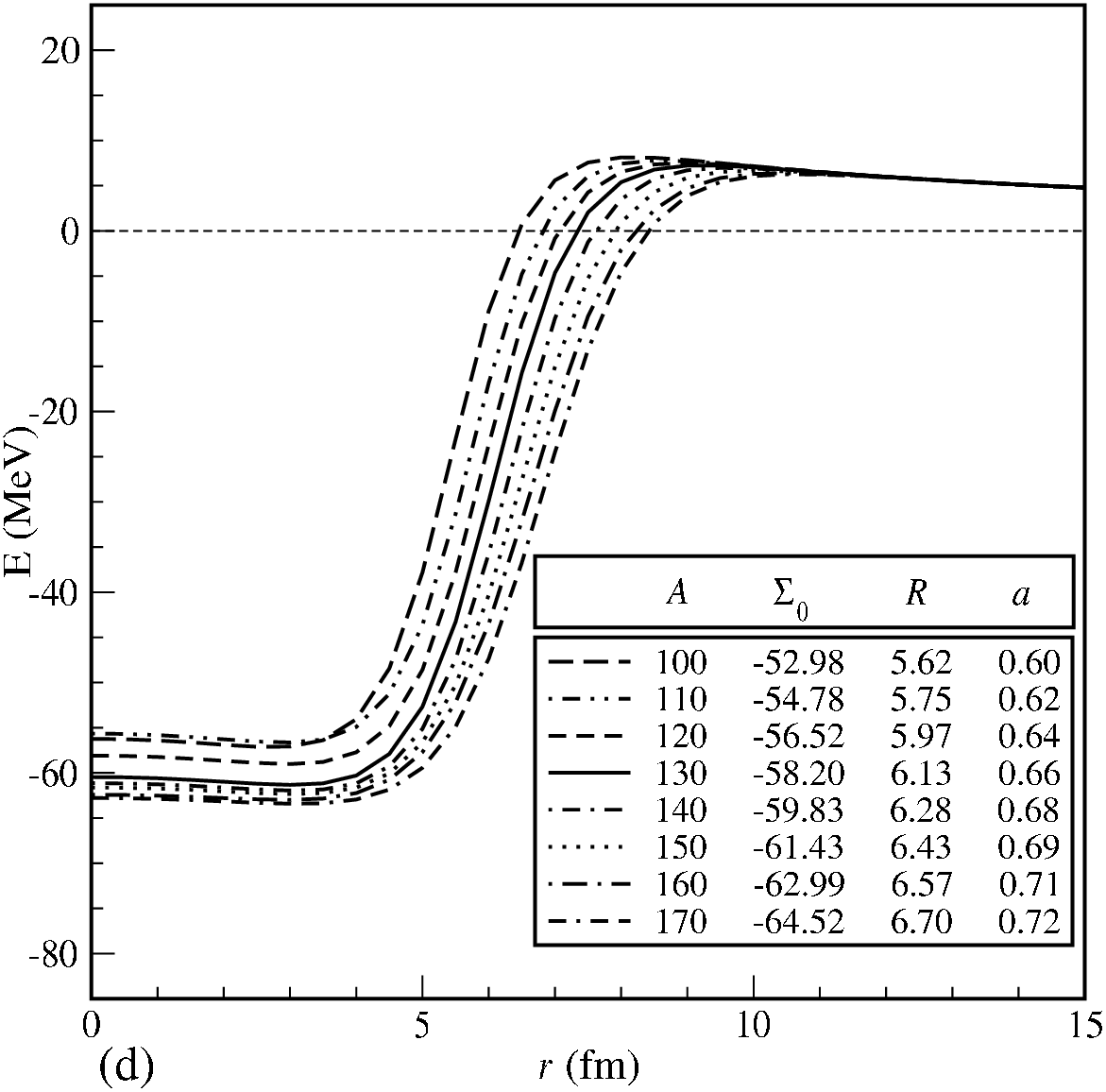}
\end{center}\vspace*{-0.4cm}}
\caption{(a) Neutron RCHB potentials for the Sn isotopic
chain from Ref.~\cite{meng}. (b) Woods-Saxon fit for the neutron
potentials in (a). The parameters $R$ and $a$ are in fermis and
$\Sigma_0$ in MeV.(c) Proton RCHB potentials for the Sn
isotopic chain from Ref.~\cite{meng}. (d)
Woods-Saxon fit for the proton potentials in (c). The units of the parameters
are as in (b).}%
\label{Fig:NeuProFit}
\end{figure}

In order to determine the effect of $V_{\rm Coul}$ and $V_\rho$ on
pseudospin splittings, we solve the Dirac equation with the
$\Sigma$ and $\Delta$ potentials for a Sn isotopic chain, from
$A=100$ to $A=170$. Since there are almost no experimental data
available for Sn neutron and proton single-particle energies, we
fit the parameters of the central potentials, $V_\rho$
[Eq.~(\ref{Vrho})] and $V_{\rm Coul}$ [Eq.~(\ref{VCoul})] to the
$\Sigma$ and $\Delta$ potentials given in Ref.~\cite{meng} for each
isotope. In that work a relativistic continuum Hartree-Bogoliubov
calculation (RCHB) with a pairing interaction, which is an
extension of the RMF theory, was performed to study tin isotopes.
As a first step to fit the neutron potential $\Sigma_n$, we used a
Woods-Saxon shape and found the corresponding parameters for each
isotope. In the case of the proton potential $\Sigma_p$, we
subtracted the Coulomb potential and fitted the nuclear part also
by a Woods-Saxon form. In order to do that, we used in
Eq.~(\ref{VCoul}) $R_{C}=1.20A^{1/3}$, which has the same $A$
dependence of the nuclear radius. In Fig.~\ref{Fig:NeuProFit} we
present the $\Sigma_n$ and $\Sigma_p$ potentials for each isotope
obtained by our fits, together with the values of the
corresponding Woods-Saxon parameters. These curves are very
similar to the ones obtained by the RCHB calculation \cite{meng}
which are also presented for comparison in Fig.~\ref{Fig:NeuProFit}.
 In the second step, using Eq.~(\ref{sigmadelta}), we extracted $\Sigma_c$ and $V_\rho$ from those
fits:
\begin{eqnarray}
\Sigma_{c}&=&\frac12(\Sigma_{p}-V_{\rm Coul}+\Sigma_{n})\label{sigma_c}\\
V_{\rho}&=&\frac12(\Sigma_{p}-V_{\rm Coul}-\Sigma_{n})\
.\label{Vr}
\end{eqnarray}
Finally, for each Sn isotope, we made a fit of $\Sigma_c$ and
$V_\rho$ matching to  Woods-Saxon shape. We found that the
parameters correlate well either with mass number $A$ or with
proton and neutron number difference $Z-N$. This allowed us to
have a general parametrization for the $\Sigma$ potential for the
whole isotopic chain:
\begin{equation}
\Sigma(r)_{p,n}=\frac{\Sigma_{o\,c}}{1+\exp[(r-R_{c})/a_{c}]}\pm\frac{V_{o\,\rho}}
{ 1+\exp[(r-R_{\rho})/a_{\rho}]} +V_{\rm Coul}(r)\label{potfull}
\end{equation}
where the last term $V_{\rm Coul}$ exists only for protons. The
parameters for  $\Sigma_{c}$ and $V_{\rho}$ potentials in
Eqs.~(\ref{sigma_c}) and (\ref{Vr}), as functions of $A$, $N$, and
$Z$, are
\begin{eqnarray}
\Sigma_{o\,c}&=&-69.94\ {\rm MeV},\\
R_{c}&=&1.21A^{1/3}\ {\rm fm},\\
a_{c}&=&0.13A^{1/3}\ {\rm fm},\\
V_{o\,\rho}&=&-[ 0.12(N-Z)+3.87]\ {\rm MeV},\label{v0rho}\\
R_{\rho}&=&[0.03(N-Z)+5.05]\ {\rm fm},\label{rrho}\\
a_{\rho}&=&[0.007(N-Z)+0.27]\ {\rm fm}.\label{arho}
\end{eqnarray}

The parametrization for $R_c$ and $a_c$ has a natural
justification in view of the known $A^{1/3}$ dependence of the
nuclear radius. The $V_{o\,\rho}$, $R_\rho$, and $a_\rho$
dependencies on $N-Z$ are also justified since they are
proportional to the difference between proton and neutron
densities. In the RCHB calculation the nuclear potentials for
protons and neutrons for the case $N=Z$ ($^{100}$Sn) are not
exactly the same. The Coulomb potential changes the proton energy
levels and because of the self-consistent RCHB calculation, the
neutron energy levels are also changed in such a way that there is
a net $V_\rho$ potential. This effect was considered in the
parametrization when we chose a value for $V_{o\,\rho}$, $R_\rho$,
and $a_\rho$ in the case of $N=Z$ as seen in 
Eqs.~(\ref{v0rho})--(\ref{arho}).

\begin{figure}[hbt]
\parbox[!t]{8.5cm}{
\begin{center}
\includegraphics[width=8.5cm,height=8.5cm]{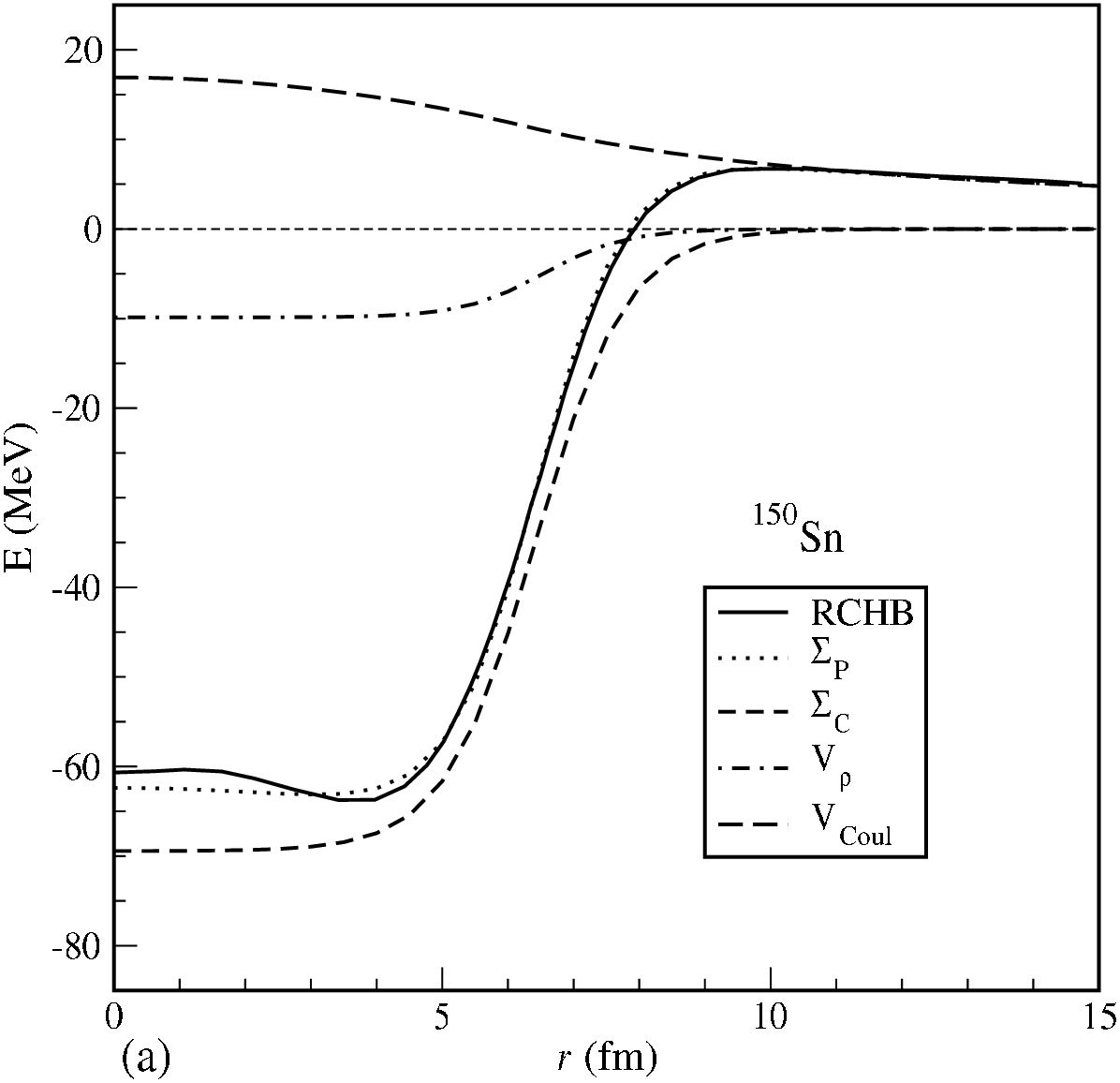}
\end{center}\vspace*{-0.4cm}
}\hfill
\parbox[!h]{8.5cm}{
\begin{center}
\includegraphics[width=8.5cm,height=8.5cm]{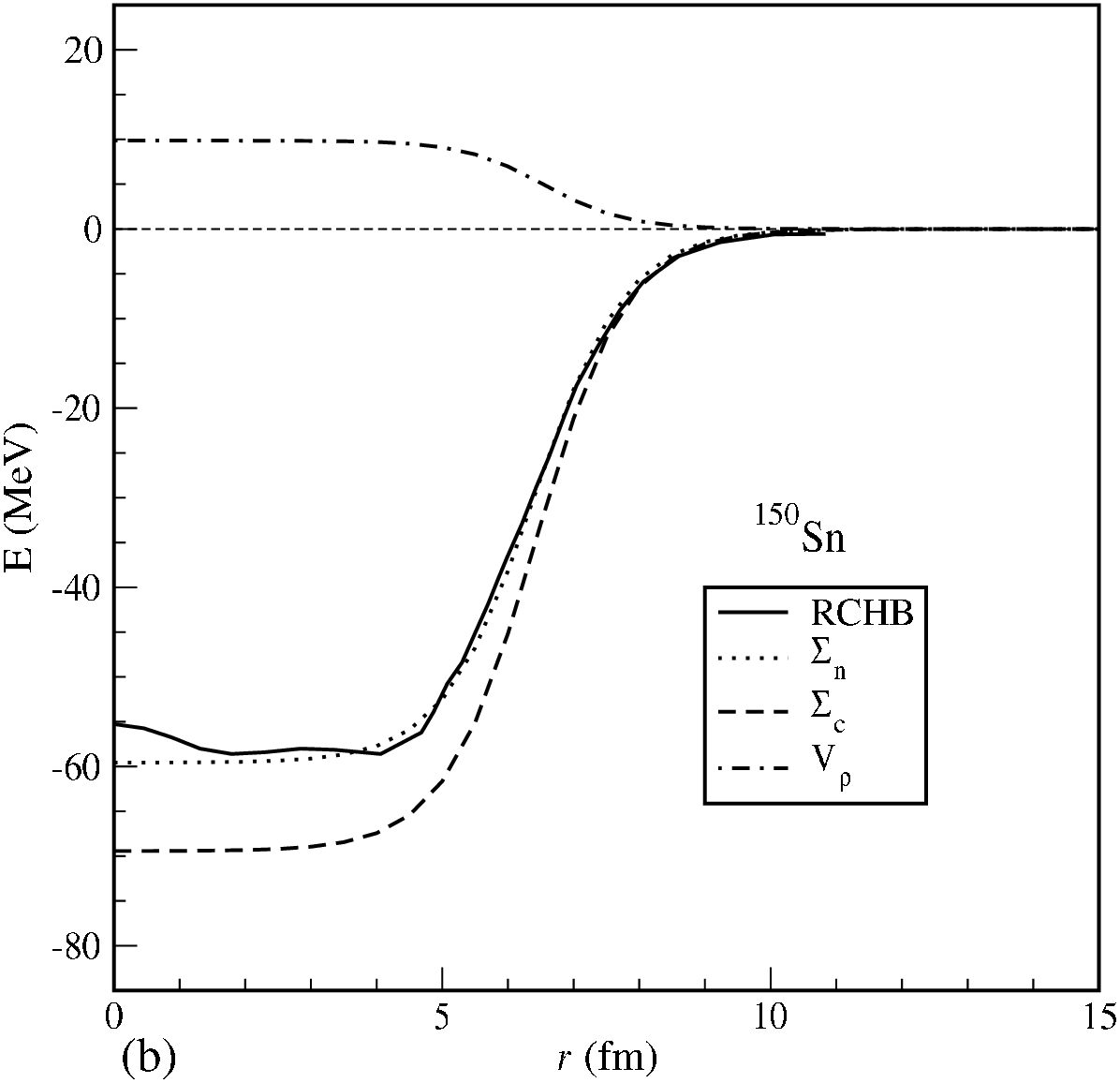}
\end{center}\vspace*{-0.4cm}
}%
\caption{(a) The contribution of $\Sigma_{c}(r)$, 
$V_{\rho}(r)$ and $V_{\rm Coul}(r)$ potentials to $\Sigma_{p}(r)$ for $^{150}$Sn. 
(b) The contribution of $\Sigma_{c}(r)$ and $V_{\rho}(r)$
potentials to $\Sigma_{n}(r)$ for $^{150}$Sn.}
\label{Fig:Sn150PN}
\end{figure}

We show in Fig.~\ref{Fig:Sn150PN} the different terms that
contribute to $\Sigma_p$ and $\Sigma_n$, respectively, for $A=150$.
In order to see the quality of our fit, we also present the
respective $\Sigma$ potentials obtained in the RCHB calculation
\cite{meng}. From Fig.~\ref{Fig:Sn150PN}(a), for the proton
potential, we see that $V_{0\,\rho}$ is around $-10$ MeV, while
$V_{\rm Coul}$ is close to $20$ MeV at the origin, so that the net
effect at the origin increases $\Sigma_c$ by $\sim 10$ MeV. In the
case of the neutron potential in Fig.~\ref{Fig:Sn150PN}(b), the net
increase at the origin turns out to be the same, because the only
contribution is from $V_{\rho}$ with a reversed sign compared to
the proton case. Taking into account the dependence of the
pseudospin splittings on the depth of the $\Sigma$ potential only,
one would expect that the proton and neutron splittings would be
almost the same. However, as we will show later, this is not the
case. The mechanism that generates the isospin asymmetry of
pseudospin is more complex.

\vspace*{0.5cm}


\section{THE ROLE OF $V_{\rm Coul}$ AND $V_\rho$ IN THE ISOSPIN ASYMMETRY OF
THE PSEUDOSPIN}

In this section, we single out the contribution of the Coulomb and
vector-isovector $V_{\rho}$ potentials to the pseudospin energy
splitting for the Sn isotopic chain. These two potentials, as we
already referred, are the main source of the isospin asymmetry of
the nuclear pseudospin: the quasidegeneracy is better seen for
neutrons than for protons.

In order to identify the origin of this asymmetry and also to see
if it has a direct relation to the pseudospin orbit potential, we
used the energy decomposition given in Eq.~(\ref{aveg_schroed}).
The proton and neutron potentials obtained for the Sn isotopes
with Woods-Saxon fit do not reproduce the small variations in the
inner part of the RCHB potentials as we can see in
Fig.~\ref{Fig:NeuProFit}. Therefore, the deeper energy levels are not
so well reproduced as the less bound ones. This should not be a
problem insofar as pseudospin symmetry in nuclei is concerned,
since it is seen experimentally that this symmetry is in general
much better satisfied for the levels near the Fermi sea.  To perform our
analysis we chose the $[2d_{5/2} - 1g_{7/2}]$
pseudospin partner, which is close to the Fermi sea. 
If we had selected any other 
doublet near the Fermi sea, the conclusion of the 
foregoing analysis would have been the same.

\begin{figure}[!h]
\parbox[!t]{8.5cm}{
\begin{center}
\includegraphics[width=8.5cm,height=8.5cm]{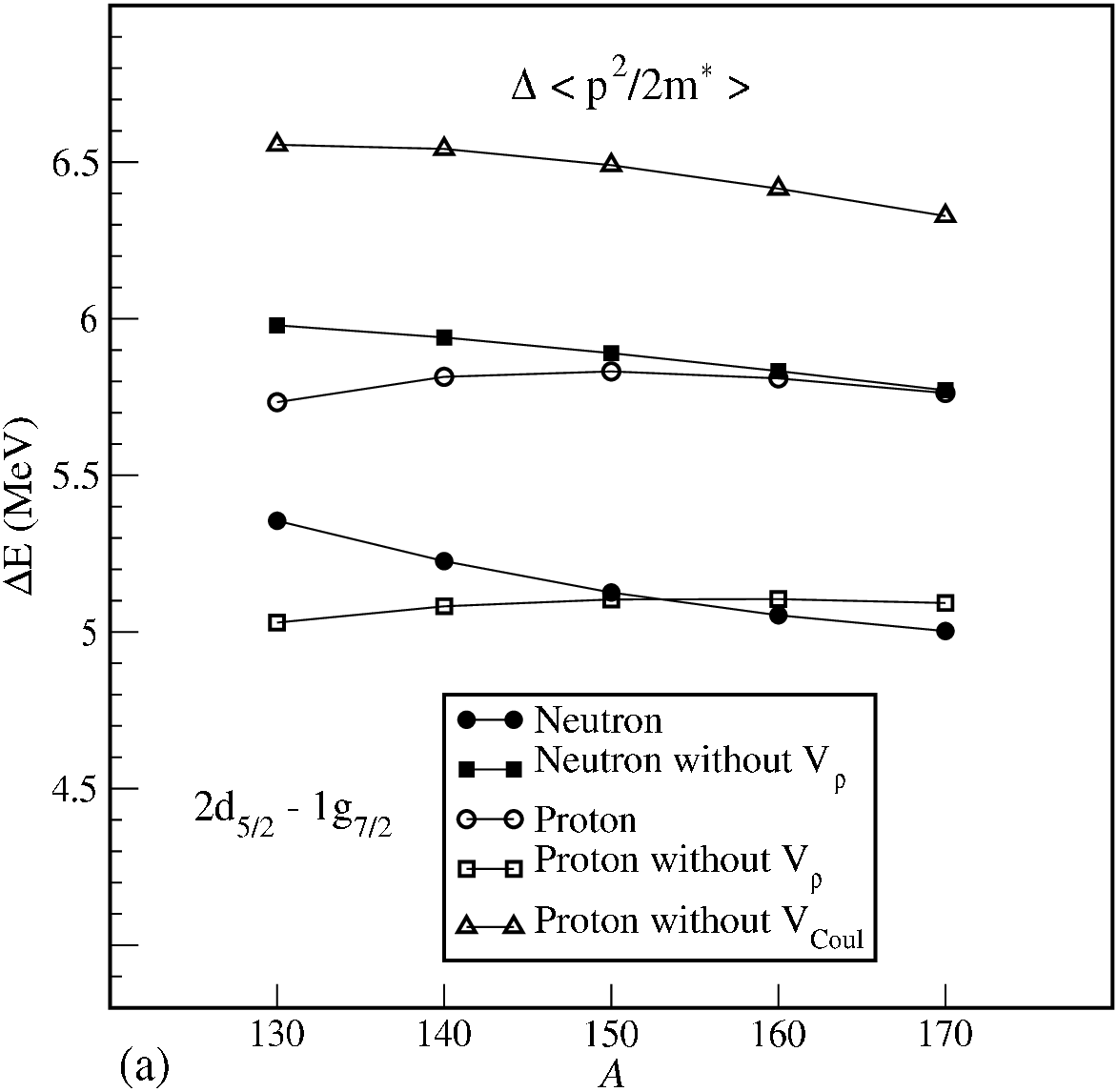}
\end{center}\vspace*{-0.4cm}}
\hfill
\parbox[!h]{8.5cm}{
\begin{center}
\includegraphics[width=8.5cm,height=8.5cm]{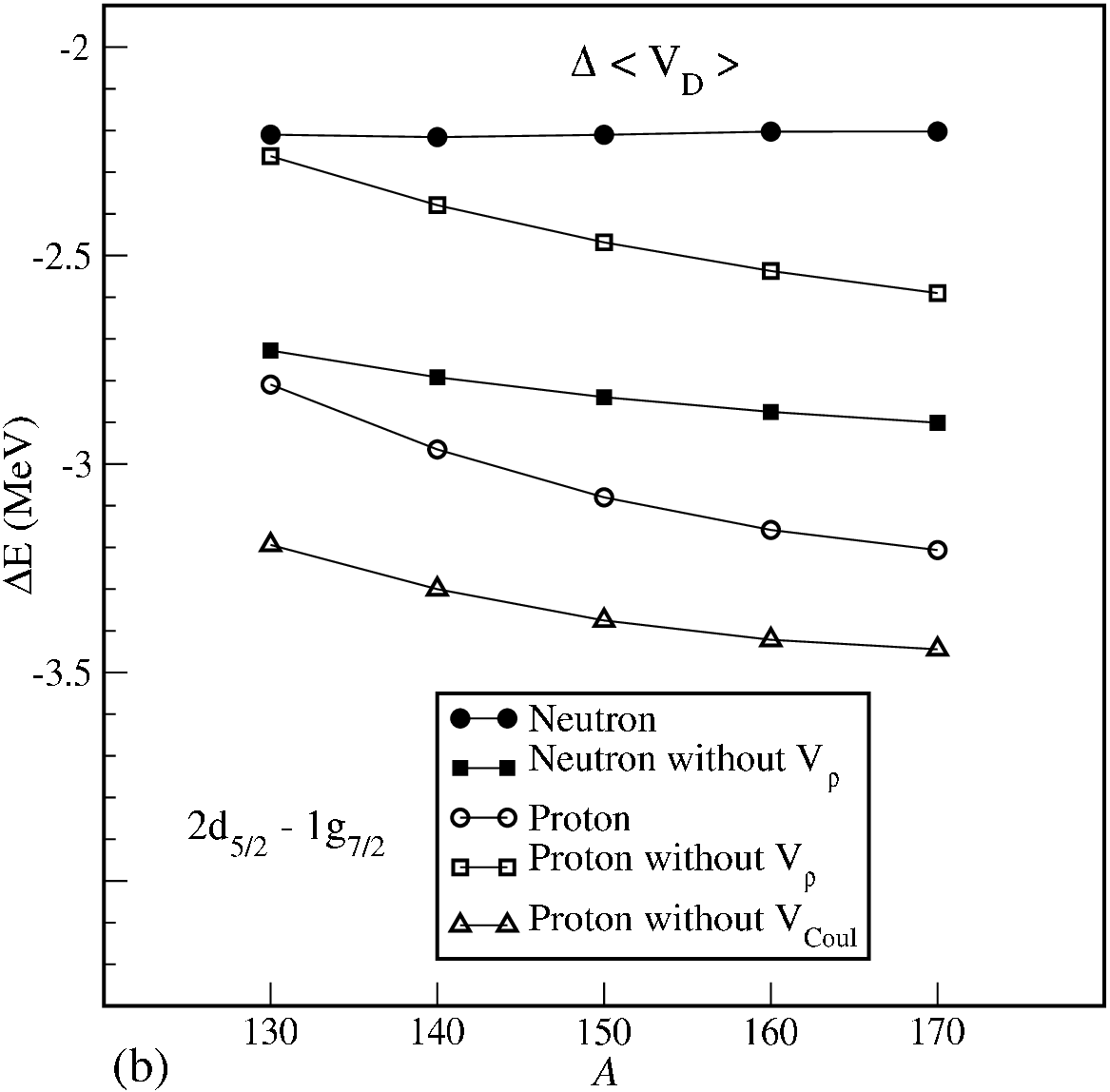}
\end{center}\vspace*{-0.4cm}}
\hfill
\parbox[!h]{8.5cm}{
\begin{center}
\includegraphics[width=8.5cm,height=8.5cm]{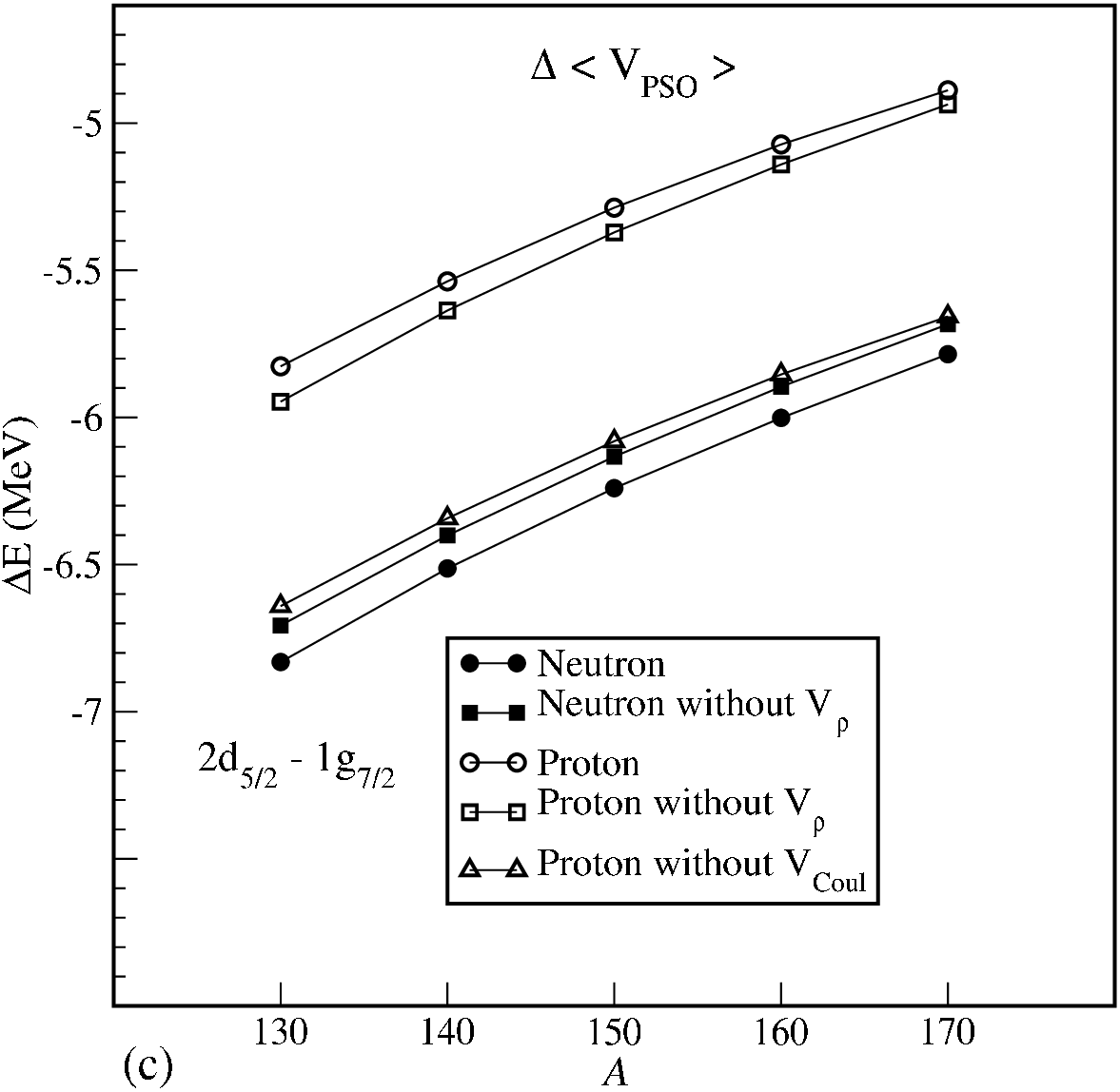}
\end{center}\vspace*{-0.4cm}}
\hfill
\parbox[!h]{8.5cm}{
\begin{center}
\includegraphics[width=8.5cm,height=8.5cm]{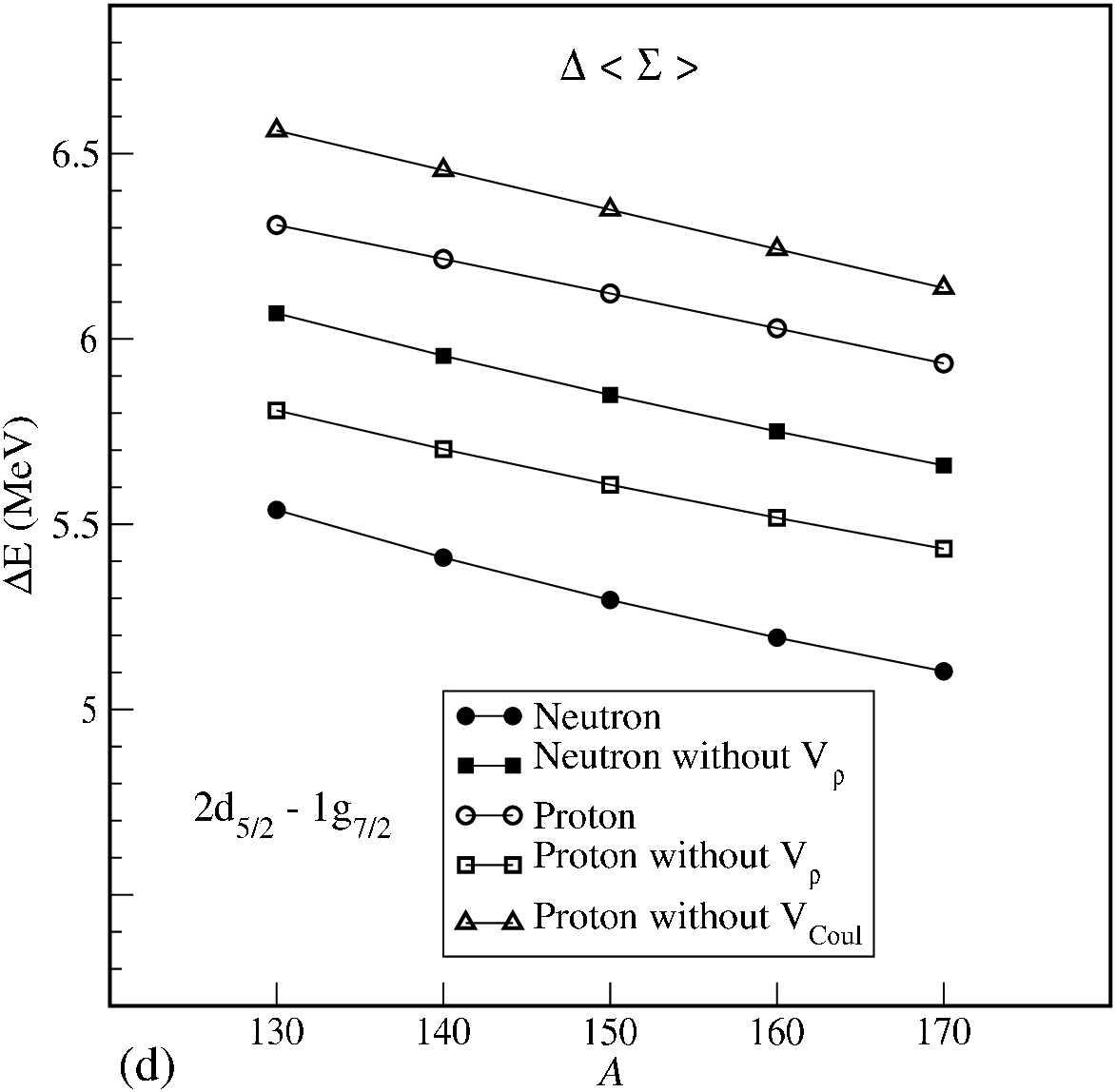}
\end{center}\vspace*{-0.4cm}}
\caption{The effect of $V_{\rm Coul}$ and $V_\rho$ in each term
of Eq.(\ref{aveg_schroed}) that contributes to the neutron and
proton pseudospin
 energy splittings for the $[2d_{5/2} -
1g_{7/2}]$ doublet in the Sn as a function of the mass number $A$.
(a) Kinetic term, (b) Darwin term, (c)
pseudospin-orbit term, (d) potential $\Sigma$
term.
} \label{Fig:Terms}
\end{figure}

\begin{figure}[hbt]
\parbox[!t]{12cm}{
\begin{center}
\includegraphics[width=8.5cm,height=8.5cm]{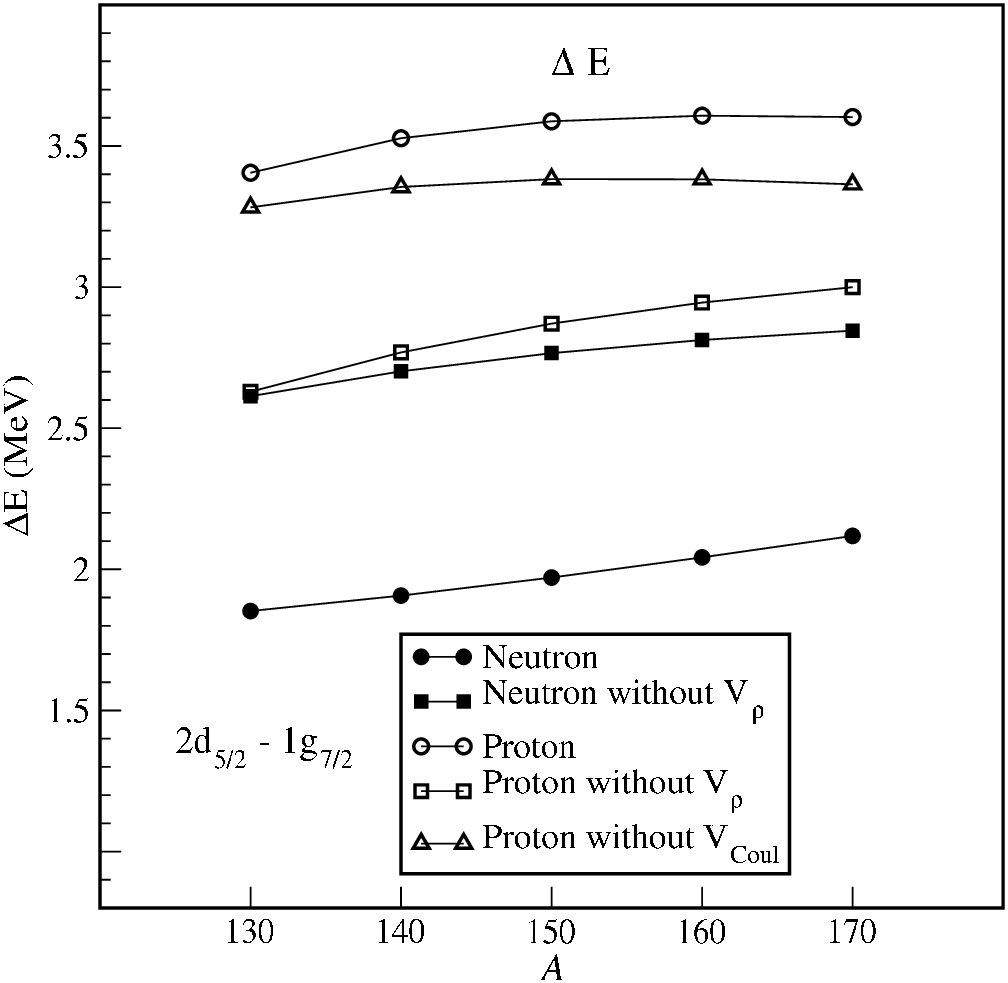}
\end{center}
\caption{The effect of $V_{\rm Coul}$ and $V_\rho$ in the neutron and
proton pseudospin energy splittings for the $[2d_{5/2} -
1g_{7/2}]$ doublet in the Sn as a function of
$A$.}\label{Fig:Terms1}}
\end{figure}

The contribution of each term of Eq.~(\ref{aveg_schroed}) to the
pseudospin energy splitting for the $[2d_{5/2} - 1g_{7/2}]$
doublet, with and without the inclusion of the Coulomb and
vector-isovector $V_{\rho}$ potentials, is shown in
Fig.~\ref{Fig:Terms} for the Sn isotopes from $A=130$ to $A=150$.
The Sn isotopes with mass number in the range $A=$100--130 were
not considered because we found that in some cases the above
levels were not bound when we excluded the $V_{\rho}$ potential.
The effect of $V_{\rm Coul}$ or $V_{\rho}$ on the energy splitting
of the pseudospin partners is presented in Fig.~\ref{Fig:Terms1}.
Let us first analyze the effect of the Coulomb potential. In
Fig.~\ref{Fig:Terms}(c) we see that the contribution from the
pseudospin orbit term $\langle V_{PS}\rangle$ to the proton energy splitting
changes considerably when we exclude the Coulomb potential. This
contribution becomes smaller ($-1$ MeV) and essentially the same
as the one for the neutron energy splitting. 
Thus, if we link the pseudospin symmetry only to
the pseudospin orbit term and do not take in account the
contribution of the other terms, we will be forced to conclude
that the Coulomb potential is the essential source of the isospin
asymmetry seen in the nuclear pseudospin. However, as it has
already been pointed out in a previous work \cite{pmmdm_prc}, the
pseudospin degeneracy comes from a significant cancellation among
all the terms in Eq.~(\ref{aveg_schroed}) and not only by the
$\langle V_{PS}\rangle$ interaction, as can be seen in 
Figs.~\ref{Fig:Terms}(a) and 
\ref{Fig:Terms}(d). 
In fact, the contribution for the proton energy difference from
kinetic $\langle p^2/2m^*\rangle$ and potential $\langle\Sigma\rangle$ energies increases
approximately the same for each of these terms ($\sim 0.5$ MeV)
when we exclude the Coulomb interaction. These effects almost
cancel the changes produced in the pseudospin orbit term
$\langle V_{PS}\rangle$. Thus, the small net effect produced by the exclusion
of the $V_{\rm Coul}$ in the proton pseudospin energy splitting
presented in Fig.~\ref{Fig:Terms1} is essentially due to the
changes in the Darwin term, which are quite small as shown in
Fig.~\ref{Fig:Terms}(b). The fact that the Coulomb potential does
not explain the isospin asymmetry reflects the dynamic nature of
the nuclear pseudospin symmetry. It comes from the contribution of
all the terms to the energy splitting and cannot be attributed
only to the pseudospin-orbit term $\langle V_{PS}\rangle$, which in this case is
strongly affected by $\langle V_{\rm Coul}\rangle$.

 Let us now analyze the effect of the vector-isovector potential
$V_{\rho}$ in this isospin asymmetry. First, we see from
Fig.~\ref{Fig:Terms}(c) that the contribution of the term $\langle V_{PS}\rangle$
to the neutron and proton energy difference remains almost
unchanged when we exclude the $V_{\rho}$ potential. The asymmetry
between protons and neutrons, as we see in Figs.~\ref{Fig:Terms}(a)
and \ref{Fig:Terms}(d), comes essentially from the effect of $V_{\rho}$ in the
kinetic and potential terms. When we exclude this potential in the
case of the neutrons, the contribution of $\langle p^{2}/2m^{\star}\rangle$ and
$\langle\Sigma\rangle$ terms are bigger (the $\Sigma$ potential is deeper),
thus increasing the neutron pseudospin energy difference as we can
see in Fig.~\ref{Fig:Terms1}. In the case of the protons we see an
opposite behavior for these two terms, since the depth of the 
$\Sigma$ potential is now smaller ($V_{\rho}$ is attractive for protons),
decreasing the proton pseudospin energy
difference. Thus, as we see in Fig.~\ref{Fig:Terms1}, the neutron
and proton pseudospin energy splittings approach each other
becoming almost degenerate. This analysis allow us to conclude
that the origin of the proton and neutron asymmetry comes mainly
from the vector-isovector $V_{\rho}$ potential.

 This conclusion is in agreement with
previous works that have also investigated this isospin asymmetry
\cite{pmmdm_prc,pmmdm_prl}. Our analysis reveals that the effect
of $V_\rho$ in the pseudospin potential in
Eq.~(\ref{aveg_schroed}) shows up mainly in an indirect way, i.e.,
through the change induced in the lower component of the wave
function $\Psi_-$. This effect has the result that, contrary to
what could be expected, the contribution of the expectation value
$\langle V_{PS}\rangle$ of the potential to the pseudospin energy splitting,
either for protons or neutrons, is almost insensitive to the
vector-isovector potential. This fact again manifests the
dynamical nature of the nuclear pseudospin symmetry. The isospin
asymmetry analyzed here is then a consequence of the  effects of
$V_{\rho}$ potential in all the terms that contribute to the
proton and neutron pseudospin energy splittings.

\section{Conclusions}

 We have investigated in a quantitative way the effect of the
Coulomb and the vector-isovector $\rho$ potentials in the
proton and neutron asymmetry seen in the nuclear pseudospin. To do
this analysis we performed a mean-field model calculation with
Woods-Saxon potentials fitted to the proton and neutron potentials
obtained by a sophisticated self-consistent RCHB calculation for
the Sn isotopes \cite{meng}. We found that the Woods-Saxon
parameters correlate well with the mass number $A$ and the proton
and neutron number difference $N-Z$, which allowed us to have a
general parametrization of the binding potential for the whole
isotopic chain. In this general parameterization of the potential
we have separated explicitly the $V_{\rm Coul}$ and $V_{\rho}$
potentials. These two potentials are the main source of the
isospin asymmetry of the nuclear pseudospin.

 In order to identify the origin of this asymmetry we analyzed the
effect of those potentials separately in the proton and
neutron pseudospin energy splitting along the Sn isotopic chain.
This analysis was done taking in account the effect of these
potentials in each of the terms that contribute to the pseudospin
energy splitting. This decomposition of the energy helped us to
uncover the mechanism behind this isospin asymmetry. We showed
that the isospin asymmetry seen in the pseudospin energy splitting
cannot be attributed to the pseudospin orbit term. It comes from a
strong cancellation among all the terms that contribute to the
energy in agreement with the dynamical nature of these symmetry
pointed out before in \cite{pmmdm_prc,pmmdm_prl,marcos,marcos2}.

 We explicitly showed that even though the pseudospin orbit term
$\langle V_{PS}\rangle$ is strongly affected by the Coulomb potential
$V_{\rm Coul}$, the net effect of this potential on the difference of
proton and neutron pseudospin energy is quite small
because of the effects of $V_{\rm Coul}$ in all the other terms.

The main conclusion of our paper is that the vector-isovector
$V_{\rho}$ potential gives the main contribution to the observed 
isospin asymmetry of the nuclear pseudospin, as stated in previous works
\cite{pmmdm_prc,pmmdm_prl}. However, it has little effect on the
contribution of the pseudospin orbit interaction term $<V_{PS}>$
to the pseudospin energy splitting, which means that the changes
of the pseudospin potential and of the wave function $F_\kappa(r)$
caused by $V_\rho$ seem to cancel each other. Therefore, the role
of $V_\rho$ in the pseudospin isospin asymmetry comes from its
effects in the other terms contributing to the single-particle
energy, especially in the differences between the kinetic and
potential energy terms of the pseudospin partners. This important
result shows that the dynamical character of pseudospin symmetry
and in particular the nonperturbative nature of the pseudospin
potential lie at the heart of the mechanism responsible for the
differences of proton and neutron pseudospin energy splittings
observed in nature.

\begin{acknowledgments}
We acknowledge financial support from the FCT (POCTI), Portugal, and
from CNPq/ICCTI Brazilian-Portuguese scientific exchange program.
R.L. acknowledges the CNPq support in particular for this work,
which is the essential part of his Master degree thesis.
\end{acknowledgments}

\end{document}